\documentclass[titlepage,
	aip,
	jcp,
	reprint,
]{revtex4-1}

\usepackage[hyperindex,breaklinks,hidelinks,colorlinks,citecolor=blue]{hyperref}
\usepackage{natbib}
\usepackage{natmove}
\usepackage{amsmath}
\usepackage{amssymb}
\usepackage{float}
\usepackage{graphicx}
\DeclareGraphicsExtensions{.pdf,.eps,.png}
\graphicspath{{./}}
\usepackage[outdir=./]{epstopdf}

\newcommand{\mt}[1]{\boldsymbol{\mathbf{#1}}}          
\newcommand{\vt}[1]{\boldsymbol{\mathbf{#1}}}          
\newcommand{\tr}[1]{#1^\text{t}}                       
\newcommand{\diff}[2]{\frac{\partial #1}{\partial #2}} 
\newcommand{\Ham}[1]{{\mathcal H}_\text{#1}}           
\newcommand{\Liu}[1]{i\!L_\text{#1}}                   
\newcommand{\timestep}{h}
\newcommand{\refined}[1]{\widetilde{#1}}
\newcommand{\mini}[1]{\text{\tiny{#1}}}

\begin{document}
	
\begin{titlepage}

\begin{flushleft}
\textbf{\large{Refinement of Thermostated Molecular Dynamics Using Backward Error Analysis}}

Ana J. Silveira and Charlles R. A. Abreu
\end{flushleft}

\textbf{This article may be downloaded for personal use only.
Any other use requires prior permission of the author and AIP Publishing.}

This article appeared in The Journal of chemical physics \textbf{150} (11), 114110 (2019) and may be found at https://aip.scitation.org/doi/abs/10.1063/1.5085441\clearpage
\end{titlepage}

\author{Ana J. Silveira}
\email{asilveira@plapiqui.edu.ar}
\affiliation{Planta Piloto de Ingenier\'ia Qu\'imica, PLAPIQUI, Universidad Nacional del Sur, Camino La Carrindanga Km 7-CC: 717, Bah\'ia Blanca, Argentina}
\altaffiliation{Current address: Computational \& Systems Biology Program, Memorial Sloan Kettering Cancer Center, New York, NY, USA}

\author{Charlles R. A. Abreu}
\email{abreu@eq.ufrj.br}
\affiliation{Chemical Engineering Department, Escola de Qu\'imica, Universidade Federal do Rio de Janeiro, Rio de Janeiro, RJ 21941-909, Brazil}

\title{Refinement of Thermostated Molecular Dynamics Using Backward Error Analysis}


\date{\today}

\begin{abstract}
Kinetic energy equipartition is a premise for many deterministic and stochastic molecular dynamics methods that aim at sampling a canonical ensemble.
While this is expected for real systems, discretization errors introduced by the numerical integration may lead to deviations from equipartition.
Fortunately, backward error analysis allows us to obtain a higher-order estimate of the quantity that is actually subject to equipartition.
This is related to a shadow Hamiltonian, which converges to the specified Hamiltonian only when the time-step size approaches zero.
This paper deals with discretization effects in a straightforward way.
With a small computational overhead, we obtain refined versions of the kinetic and potential energies, whose sum is a suitable estimator of the shadow Hamiltonian.
Then, we tune the thermostatting procedure by employing the refined kinetic energy instead of the conventional one.
This procedure is shown to reproduce a canonical ensemble compatible with the refined system, as opposed to the original one, but canonical averages regarding the latter can be easily recovered by reweighting.
Water, modeled as a rigid body, is an excellent test case for our proposal because its numerical stability extends up to time steps large enough to yield pronounced discretization errors in Verlet-type integrators.
By applying our new approach, we were able to mitigate discretization effects in equilibrium properties of liquid water for time-step sizes up to 5 fs.
\end{abstract}

\maketitle

\section{Introduction}

Molecular dynamics (MD) simulation involves several types of approximations in the quest for mimicking real systems.
Naively, we tend to rank simulation results by their proximity to experimental data and attribute any disagreement to a lack of accuracy of the employed force field.
However, systematic errors can possibly be hidden in these deviations, being the cause of non-physical behavior.
It has been recently shown \cite{Palmer_2018} that erroneous simulation results \cite{Limmer_2011, Limmer_2013} were the cause of a long-standing controversy \cite{Smart_2018} about a liquid-liquid transition believed to exist for supercooled water.
In this case, hybrid Monte Carlo steps were employed with initial velocities that failed to comply with the equipartition between translational and rotational degrees of freedom \cite{Palmer_2018}.
As a matter of fact, similar ill-partitioning of kinetic energy can arise in MD simulations, either associated with noncanonical rescaling of velocities \cite{Braun_2018} or as an artifact due to the discretization of the equations of motion.
This can affect not only the translation/rotation equipartition in rigid-body systems \cite{Davidchack_2010, Silveira_2017}, but also the distribution of energy among molecules of distinct sizes \cite{Eastwood_2010}.
The issue with rigid-body MD is relevant not only because water \cite{Jorgensen_1983} and other small molecules are usually modeled as rigid bodies, but also because designing proteins, nanoparticles, or other large structures as collections of interconnected bodies \cite{Miller_2002, Knorowski_2012, Patra_2013} can be seen as a useful coarse-graining strategy.
It also brings into question the NVT-MD method developed by Kamberaj, Low, and Neal \cite{Kamberaj_2005} (KLN) who, by assuming equipartition, coupled two independent thermostats to the translational and rotational degrees of freedom.
As we will show here, the method fails to reproduce the specified temperature even for time-step sizes typically employed in MD.
In fact, Davidchack \cite{Davidchack_2010} recommends that the use of separate thermostats for translation and rotation should always be avoided.
Nonetheless, the cited double-thermostat method is available in software packages such as HOOMD-\textit{blue}\cite{Anderson_2008} and LAMMPS\cite{Plimpton_1995} and has been applied to simulate small molecules \cite{Geiger_2013, Aimoli_2014, Aimoli_2014_2}, membranes \cite{Bucior_2012}, molecular motors \cite{Akimov_2012}, micelles \cite{Yan_2008}, and nanoparticles \cite{Patra_2014}, in spite of all the issues observed here.

For a symplectic integration of Hamiltonian equations of motion, the breakdown of equipartition is arguably only apparent \cite{Eastwood_2010}.
The produced trajectory actually follows the dynamics dictated by a \textit{shadow Hamiltonian} \cite{Tuckerman_2010}, which is no longer separable into exclusively momentum- and position-dependent parts and only coincides with the specified Hamiltonian as the time-step size approaches zero.
Therefore, what seems to occur, in fact, is a failure to identify the quantity that is subject to equipartition in place of the specified kinetic energy.
As Eastwood \textit{et al}. \cite{Eastwood_2010} have shown, Tolman's generalized equipartition theorem \cite{Tolman_1918} holds at least for small steps, when such a \textit{shadow kinetic energy} keeps displaying a quadratic dependency on the momenta.
The authors then employed backward error analysis to successfully eliminate the artifacts mentioned above.
As we manage to demonstrate here, their findings can be extended to the symplectic integration of rigid-body motion as well.

Issues arise in NVT-MD because thermostatting methods usually rely on the corollary of the equipartition theorem which relates temperature to kinetic energy.
Thus, it seems reasonable to argue that a thermostat should regulate the shadow kinetic energy instead of the specified one.
However, the theory underlying the concept and computation of shadow Hamiltonians is only applicable to symplectic equations and symplectic integrators, thus ruling out most NVT-MD methods in use today.
In this paper, we devise a novel procedure that is able to overcome this limitation in the case of global thermostatting methods such as, for instance, the Nos\'{e}-Hoover Chain (NHC) \cite{Martyna_1992}, the Stochastic Velocity Rescaling \cite{Bussi_2007}, and the Nos\'{e}-Hoover-Langevin \cite{Samoletov_2007, Leimkuhler_2009} methods.
It consists in using a splitting-based numerical integrator that entails a Hamiltonian component and then applying backward error analysis to that part only.
This allows one to obtain, with minimal computational overhead, a \textit{refined kinetic energy} and a \textit{refined Hamiltonian} as suitable estimators of their shadow counterparts.
By recasting the thermostat equations with such a refined kinetic energy, we are able to predict its equipartition and reproduce a canonical ensemble consistent with the refined Hamiltonian.
Finally, a reweighting procedure yields ensemble averages which are, instead, consistent with the specified one.
For simplicity, we restrict our detailed derivation to the case of a NHC thermostat \cite{Martyna_1992}.

The new method is especially appealing for rigid-body MD, primarily because numerical stability can, in this case, extend up to time-step sizes large enough to yield pronounced discretization errors.
For rigid water, which served as an excellent test case, we were able to correct discretization errors in ensemble averages computed with time-step sizes up to $5~\text{fs}$, thus outperforming not only the KLN method \cite{Kamberaj_2005}, but also a reimplementation for rigid bodies of the NHC integrator introduced by Martyna \textit{et al}. \cite{Martyna_1996}, as well as stochastic rescaling \cite{Bussi_2007} applied simultaneously to translational and angular velocities.

This paper proceeds as follows.
In Secs.~\ref{sec:nve} and \ref{sec:nvt} we review the rigid-body dynamics in the NVE and NVT ensembles.
This is followed by our developments on backward error analysis, which we employ in Sec.~\ref{sec:refined_method} to derive a new NVT scheme, whose performance we assess in Sec.~\ref{sec:numerical_results}.
Finally, we present some concluding remarks in Sec.~\ref{sec:conclusion}. Appendix \ref{sec:rigid body shadow hamiltonian} contains a detailed derivation of the shadow Hamiltonian approximation for a system containing rigid bodies.

\section{Methodology}
\label{sec:methodology}

\subsection{Hamiltonian Dynamics with Rigid Bodies: Notation and Symplectic Integration}
\label{sec:nve}

In a previous paper \cite{Silveira_2017}, we took advantage of a particular factorization of the orientation matrix expressed in terms of quaternion components to derive a new exact solution for torque-free rotations and use it as part of a symplectic integrator for rigid bodies.

The Hamiltonian system of ordinary differential equations (ODE) that describes the rigid-body motion is given by \cite{Silveira_2017}
\begin{subequations}
	\label{eq:ODE system for NVE}
	\begin{align}
	&\dot{\vt r} =
	{\mt M}^{-1} {\vt p}, \\
	&\dot{\vt p} =
	{\vt F}, \\
	&\dot{\vt q} =
	\frac{1}{2} \mt B(\vt q) \vt \omega, \text{ and} \label{eq:EDO_q} \\
	&\dot{\vt \pi} =
	\frac{1}{2} \mt B(\vt \pi) \vt \omega + 2 \mt C(\vt q) \vt \tau. \label{eq:EDO_pi}
	\end{align}
\end{subequations}

In these equations, $\vt r$ is the center-of-mass position of the body, $\vt p$ is its linear momentum, $\vt q$ is a unit quaternion that determines its orientation, and $\vt \pi$ is the quaternion-conjugate momentum.
Vectors $\vt F$ and $\vt \tau$ are, respectively, the resultant force and torque exerted on the body, both represented in the space-fixed frame of reference, and $\vt \omega = \frac{1}{2} {\mt I}^{-1} \tr{\mt B}(\vt q) {\vt \pi}$ is the three-dimensional angular velocity in the body-fixed frame.
Matrices $\mt M$ and $\mt I$ are diagonal ones.
The former contains the body mass in all three diagonal entries while the latter contains the three principal moments of inertia.
Finally, the matrix-valued functions $\mt B$ and $\vt C$, given by
\begin{equation*}
\label{eq:def_B_and_C}
\mt B(\vt q) = \left[\begin{array}{rrrr}
-q_2 & -q_3 & -q_4 \\
 q_1 & -q_4 &  q_3 \\
 q_4 &  q_1 & -q_2 \\
-q_3 &  q_2 &  q_1
\end{array}\right]
\end{equation*}
and
\begin{equation*}
\mt C(\vt q) = \left[\begin{array}{rrrr}
-q_2 & -q_3 & -q_4 \\
 q_1 &  q_4 & -q_3 \\
-q_4 &  q_1 &  q_2 \\
 q_3 & -q_2 &  q_1
\end{array}\right]
\end{equation*}
are related to the rotation matrix ${\mt A}(\vt q)$ through the factorization ${\mt A}(\vt q) = \tr{\mt B}(\vt q) {\mt C}(\vt q)$ that was mentioned at the beginning of this section. Eq.~\eqref{eq:ODE system for NVE} preserves the constraints $\tr{\vt q}{\vt q} = 1$ and $\tr{\vt q}{\vt \pi} = 0$, as well as the value of a Hamiltonian
\begin{equation}
\label{eq:Hamiltonian}
\Ham{} = K(\vt p, \vt q, \vt \pi) + U(\vt r,\vt q),
\end{equation}
where $K$ and $U$ are the kinetic and potential energies of the body, respectively.
While the form of $U(\vt r, \vt q)$ depends on the specific interaction model, the kinetic energy is given by
\begin{equation}
\label{eq:kinetic energy}
K = \frac{1}{2} \tr{\vt p} {\mt M}^{-1} {\vt p} + \frac{1}{8} \tr{\vt \pi} {\mt B}(\vt q) {\mt I}^{-1} \tr{\mt B}(\vt q) {\vt \pi}.
\end{equation}

Eqs.~\eqref{eq:ODE system for NVE} to \eqref{eq:kinetic energy} can also represent a system of $N$ interacting rigid bodies if we consider that vectors $\vt r$, $\vt p$, $\vt q$, and $\vt \pi$ contain the corresponding properties of all bodies combined in a single vector.
The size of diagonal matrices $\mt M$ and $\mt I$ also increases to $3N \times 3N$ in this case.
In addition, $\mt B(\vt q)$ and $\mt C(\vt q)$ become block-diagonal matrices with $4N$ rows and $3N$ columns.
Yet, the notation can be made even more general if $\vt r$, $\vt p$, and $\mt M$ are enlarged further so as to accommodate a number of individual point masses.

A numerical solution of Eq.~\eqref{eq:ODE system for NVE} is usually represented by a stepwise application of the classical time-evolution propagator \cite{Tuckerman_2010} $e^{\timestep \Liu{\tiny NVE}}$ to the system configuration, where $\timestep$ is the time-step size and $\Liu{\tiny NVE}$ is the Liouville operator associated to $\Ham{}$, which is
\begin{equation}
\label{eq:Lioville operator}
\begin{split}
\Liu{\tiny NVE} &= \tr{\vt p} {\mt M}^{-1} \diff{}{\vt r} + \frac{1}{2} \tr{\vt \omega} \tr{\mt B}(\vt q) \diff{}{\vt q} + \tr{\vt F} \diff{}{\vt p} \\
&+ \left[ \frac{1}{2} \tr{\vt \omega} \tr{\mt B}(\vt \pi) + 2 \tr{\vt \tau} \tr{\mt C}(\vt q) \right] \diff{}{\vt \pi}.
\end{split}
\end{equation}

A symplectic, time-reversible integrator can be devised by splitting the exponential operator according to the Trotter-Suzuki formula \cite{Trotter_1959, Suzuki_1976}
\begin{equation}
\label{eq:splitting NVE}
e^{\timestep \Liu{\tiny NVE}} = e^{\frac{\timestep}{2} \Liu{B}} e^{\timestep \Liu{A}} e^{\frac{\timestep}{2} \Liu{B}} + \mathcal{O}(\timestep^3),
\end{equation}
where $\Liu{A} + \Liu{B} = \Liu{\tiny NVE}$.
In the Verlet-type splitting of Ref.~\citenum{Silveira_2017}, the action of propagator $e^{\frac{\timestep}{2} \Liu{B}}$ is a kick that changes the linear and quaternion momenta according to ${\vt p} = {\vt p}^\ast + \frac{\timestep}{2} {\vt F}^\ast$ and ${\vt \pi} = {\vt \pi}^\ast + \timestep {\mt C}({\vt q^\ast}) {\vt \tau}^\ast$, respectively; where the asterisk denotes the state of the system immediately before the propagation occurs.
The action of $e^{\timestep \Liu{A}}$ is a simultaneous uniform translation (${\vt r} = {\vt r}^\ast + \timestep {\mt M}^{-1} {\vt p}^\ast$) and torque-free rotation along a full time step.
We have derived \cite{Silveira_2017} an exact solution for torque-free rotations in a form that facilitates computer implementation and allows, differently from other existing solutions \cite{Kosenko_1998, vanZon2007, Celledoni_2008}, a unified treatment of asymmetric, symmetric, and spherical tops.
This solution was used to evaluate the simpler NO-SQUISH method \cite{Miller_2002}.
Despite its approximate treatment of rotations, which are split into several revolutions around the principal axes \cite{Dullweber_1997}, that method proved to be very accurate in liquid-phase simulations \cite{Silveira_2017}.
This justifies its more widespread use.
However, having an exact solution for free rotations at hand is crucial for the feasibility of backward error analysis, as it will become clear in Sec.~\ref{sec:backward error analysis} and Appendix \ref{sec:rigid body shadow hamiltonian}.

\subsection{NVT Dynamics with a Single Nos\'{e}-Hoover Chain: Notation and Measure-Preserving Integration}
\label{sec:nvt}

In this work, we couple a single Nos\'{e}-Hoover chain thermostat \cite{Martyna_1992} to both the translational and rotational degrees of freedom of the rigid-body system.
To this end, we consider an extra generalized coordinate $\eta_j$ and its conjugate momentum $p_{\eta_j}$ for each thermostat $j = 1, \cdots, M$.
The flow in the extended phase-space no longer conserves the Hamiltonian $\Ham{}$, but an \textit{extended energy} given by \cite{Martyna_1992}
\begin{equation}
\label{eq:nvt extended energy}
H = K + U + {\textstyle\sum\limits_{j=1}^{M}} \frac{p_{\eta_j}^2}{2Q_j} + L k_\mini{B} T\eta_1 + k_\mini{B} T {\textstyle\sum\limits_{j=2}^{M}} \eta_j,
\end{equation}
where $k_\mini{B}$ is the Boltzmann constant, $T$ is the target temperature, $L$ is a constant to be determined, and each $Q_j$ is an inertial parameter.
As recommended in Ref.~\citenum{Martyna_1992}, we can make $Q_1 = L k_\mini{B} T t_d^2$ and $Q_j = k_\mini{B} T t_d^2$ for $j \geq 2$, where $t_d$ is a characteristic time scale of the thermostat chain.

By employing the method of Sergi and Ferrario \cite{Sergi_2001}, we obtain the equations of motion for the NVT dynamics with a single thermostat chain, which are
\begin{subequations}
	\label{eq:ODE system for NVT}
	\begin{align}
\label{eq:nhc_r}
	&\dot{\vt r} =
	{\mt M}^{-1} {\vt p}, \\
\label{eq:nhc_p} 
	&\dot{\vt p} =
	{\vt F} - \alpha_1 {\vt p},\\
\label{eq:nhc_q}
	&\dot{\vt q} =
	\frac{1}{2} \mt B(\vt q) \vt \omega, \\
\label{eq:nhc_pi}
	&\dot{\vt \pi} =
	\frac{1}{2} \mt B(\vt \pi) \vt \omega + 2 \mt C(\vt q) \vt \tau - \alpha_1 {\vt \pi}, \\
\label{eq:nhc_eta}
	&\dot{\eta}_j = \alpha_j, \text{ and} \\
\label{eq:nhc_p_eta}
	&{\dot p}_{\eta_j} = G_j - \alpha_{j+1} p_{\eta_j} \qquad \text{for} \quad 1 \leq j \le M.
	\end{align}
\end{subequations}

In these equations, $\alpha_j$ $=$ ${p_{\eta_j}}/{Q_j}$ for $j \le M$ and $\alpha_{M+1} = 0$, while $G_j$ is a generalized force acting on thermostat $j$, defined as
\begin{subequations}
\begin{align}
\label{eq:generalized force NVT}
&G_1 = \tr{\vt p} \diff{K}{\vt p} + \tr{\vt \pi} \diff{K}{\vt \pi} - L k_\mini{B} T \quad \text{and}\\
&G_j = \frac{p_{\eta_{j-1}}^2}{Q_{j-1}} - k_\mini{B} T  \qquad \text{for} \quad j > 1.
\end{align}
\end{subequations}

With $K$ defined as in Eq.~\eqref{eq:kinetic energy}, it turns out that $G_1 = 2K - L k_\mini{B} T$.
If there are no external forces, the vector quantity $e^{\eta_1}\sum_{i=1}^N {\vt p}_i$ and the extended energy $H$ are conserved, while the system's center-of-mass position and the $\eta$ coordinates can be considered as driven variables \cite{Tuckerman_2001}.
Moreover, $2N$ equations must be eliminated due to the constraints involving $\vt q$ and $\vt \pi$.
In the general case, by taking all these facts into account and applying the analysis of Tuckerman \textit{et al}. \cite{Tuckerman_2001}, we can deduce that the correct canonical distribution is attained if we make $L = 6N$.
In a particular situation in which we set $\sum_{i=1}^N {\vt p}_i = \vt 0$ at the onset of the simulation, we must make $L = 6N - 3$ \cite{Martyna_1994}.

Eq.~\eqref{eq:ODE system for NVT} defines an invariant measure in the extended phase-space \cite{Tuckerman_1999}, and it is possible to devise numerical solvers that preserve such measure \cite{Sergi_2001, Ezra_2004, Ezra_2006}.
We have done it by adapting the integrator introduced by Martyna \textit{et al}. \cite{Martyna_1996} in light of a criterion explained by Ezra \cite{Ezra_2006}.
It is worth noting, however, that such adaptation required us to only alter the integration of the $\eta$ coordinates.
Since these are driven variables, with no influence on the dynamics of the system, we can conclude that the practical importance of such alteration is small.

By defining a propagator $e^{\timestep \mathcal{L}_\mini{NVT}}$, where $\mathcal{L}_\mini{NVT}$ is a generalized (i.e. non-Hamiltonian) Liouville operator, the splitting goes as
\begin{equation}
\label{eq:splitting NVT}
e^{\timestep \mathcal{L}_\mini{NVT}} = e^{\frac{\timestep}{2} \mathcal{L}_\mini{NHC}} e^{\timestep \Liu{\tiny NVE}} e^{\frac{\timestep}{2} \mathcal{L}_\mini{NHC}} + \mathcal{O}(\timestep^3),
\end{equation}
where $\Liu{\tiny NVE}$ is the same operator of Sec.~\ref{sec:nve} and $\mathcal{L}_\mini{NHC}$ aggregates all new terms introduced by the Nos\'e-Hoover chain.
This one can be split even further as
\begin{equation*}
e^{\frac{\timestep}{2} \mathcal{L}_\text{NHC}} = \left[ \left( \textstyle\prod\limits_{j=M}^1 e^{\frac{\timestep}{4n} \mathcal{L}_j }\right) e^{\frac{\timestep}{2n} \mathcal{L}_0 } \left(  \textstyle\prod\limits_{j=1}^M e^{\frac{\timestep}{4n} \mathcal{L}_j }\right)  \right]^n.
\end{equation*}

In the scheme above, the first propagator to act is $e^{\frac{\timestep}{4n} \mathcal{L}_M}$, which promotes a kick in the momentum of thermostat $M$ expressed as $p_{\eta_M} = p_{\eta_M}^\ast + G_M^\ast \frac{\timestep}{4n}$.
Then, propagators $e^{\frac{\timestep}{4n} \mathcal{L}_j}$ act sequentially, with $j$ varying from $M-1$ down to $1$.
With $\phi(x) = \frac{1-e^{-x}}{x}$, the effect of each one is translated as \cite{Martyna_1996}
\begin{align*}
&p_{\eta_j} = p_{\eta_j}^\ast + \Big( G_j^\ast - \alpha_{j+1}^\ast p_{\eta_j}^\ast \Big) \phi\left(\frac{\alpha_{j+1}^\ast \timestep}{4n}\right) \frac{\timestep}{4n} \; \text{and} \\
&\eta_{j+1} = \eta_{j+1}^\ast + \alpha_{j+1}^\ast \frac{\timestep}{4n}.
\end{align*}

After that, propagator $e^{\frac{\timestep}{2n} \mathcal{L}_0}$ establishes the effect of thermostat $1$ on the motion of the rigid bodies, as well as the evolution of coordinate $\eta_1$.
Its action is expressed as ${\vt p} = e^{-\alpha_1^\ast \frac{\timestep}{2n}} {\vt p}^\ast$, ${\vt \pi} = e^{-\alpha_1^\ast \frac{\timestep}{2n}} {\vt \pi}^\ast$, and $\eta_1 = \eta_1^\ast + \alpha_1^\ast \frac{\timestep}{2n}$.
Finally, propagators $e^{\frac{\timestep}{4n} \mathcal{L}_j}$ are applied once again, but now with index $j$ ascending from $1$ to $M$.

Due to its low computational cost, the whole procedure described above can be repeated $n$ times with little impact on the overall integration effort.
This makes it feasible to increase the time step $\timestep$ without ruining the accuracy of the NHC part.
In contrast, evaluating the NVE part is expensive for involving force/torque computations and, in addition, its accuracy goes down quickly as $\timestep$ increases \cite{Davidchack_2010, Silveira_2017}.
A high-order integration scheme\cite{Omelyan_2007, Van_zon_2008} could possibly admit larger time steps, but it would raise both cost and complexity due to the need to evaluate (or approximate numerically) the force/torque gradients.
Omelyan's processed splitting approach \cite{Omelyan_2008} seems promising, but its extension to the NVT case without doing back-and-forth processing at every time step would require further theoretical development.
Here we choose to take a different path.
Instead of trying to increase accuracy in the evaluation of $e^{\timestep \Liu{\tiny NVE}}$, we attempt to quantify the discretization errors, with which we can both tune the thermostatting procedure and properly weight the sampled configurations when ensemble averages are computed.

\subsection{Backward Error Analysis}
\label{sec:backward error analysis}

In order to explain our approach, we turn the attention again to the NVE case.
A known property of splitting methods applied to Hamiltonian ODE systems is that they provide approximate solutions which are Hamiltonian as well.
Hence, a trajectory generated to be roughly consistent with the specified function $\Ham{}$ will be exactly consistent with a nearby (albeit unknown) function $\Ham{S}$, referred to as a shadow Hamiltonian \cite{Hairer_2006}, which explicitly depends on the time-step size $\timestep$.

A new result we present here is an analytically derived approximation for $\Ham{S}$, referred to here as a \textit{refined Hamiltonian} $\refined{\Ham{}}$, concerning a system of rigid bodies whose dynamics is integrated via the unsplit rotation method of Ref.~\citenum{Silveira_2017}.
A detailed derivation is provided in Appendix~\ref{sec:rigid body shadow hamiltonian}.
As demonstrated therein, such approximation is given by $\Ham{S} = \refined K + \refined U + \mathcal{O}(\timestep^4)$, where $\refined K$ and $\refined U$ are the refined versions of the kinetic and potential energies, respectively.
The latter is given by
\begin{equation}
\label{eq:modified potential energy}
\refined U = U - \frac{\timestep^2}{24} \left( \tr{\vt F} {\mt M}^{-1} {\vt F} + \tr{\vt \tau} \tr{\mt A} {\mt I}^{-1} {\mt A} {\vt \tau} \right).
\end{equation}

Note that $\refined U$ becomes dependent on $\timestep^2$ but, like $U$, it is independent of $\vt p$ and $\vt \pi$.
In turn, the \textit{refined kinetic energy} can be expressed as
\begin{equation}
\label{eq:modified kinetic energy}
\refined K = \frac{1}{2} \tr{\left[\begin{array}{c} \vt p \\ \vt \pi \end{array}\right]} \refined{\mathbf \Omega}(\vt r, \vt q, \timestep^2) \left[\begin{array}{c} \vt p \\ \vt \pi \end{array}\right],
\end{equation}
where $\refined{\mathbf \Omega}$ is a matrix-valued function of $\vt r$, $\vt q$, and $\timestep^2$ given by $\tilde{\mt \Omega} = {\mt \Omega} + ({\timestep^2}/{6}) {\mt \Omega} {\mt \Xi} \tr{\mt \Omega}$.
In this definition, $\mt \Xi$ is the Hessian matrix of the potential energy with respect to $[\substack{\vt r \\ \vt q}]$, while ${\mt \Omega}$ is a block-diagonal matrix defined so that $K = \frac{1}{2} \tr{[\substack{\vt p \\ \vt \pi}]} {\mt \Omega} [\substack{\vt p \\ \vt \pi}]$.
Although $\refined{\mathbf \Omega}$ is symmetric like $\mt \Omega$, its structure is not block-diagonal due to the inter-body interactions accounted for in $U(\vt r, \vt q)$ and to the dependency of forces and torques on both $\vt r$ and $\vt q$.
As a result, one cannot generally split the refined kinetic energy into independent translational and rotational contributions.

Now notice that a trajectory generated by the splitting scheme of Eq.~\eqref{eq:splitting NVE}, which is intended to approach the exact solution of Eq.~\eqref{eq:ODE system for NVE}, will even more closely approach the solution of a modified ODE system given by
\begin{equation*}
\dot{\vt r} = \diff{\refined K}{\vt p}, \;
\dot{\vt q} = \diff{\refined K}{\vt \pi}, \;
\dot{\vt p} = -\diff{\refined{\Ham{}}}{\vt r}, \; \text{and} \;
\dot{\vt \pi} = -\diff{\refined{\Ham{}}}{\vt q}.
\end{equation*}

The first two of these equations can be combined together so that we can use Eq.~\eqref{eq:modified kinetic energy} to obtain
\begin{equation}
\label{eq:shadow ODE system}
\left[\begin{array}{c} \dot{\vt r} \\ \dot{\vt q} \end{array}\right] = \refined{\mathbf \Omega}(\vt r, \vt q, \timestep^2) \left[\begin{array}{c} \vt p \\ \vt \pi \end{array}\right].
\end{equation}

Analytical calculation of $\refined U$ via Eq.~\eqref{eq:modified potential energy} is straightforward.
In contrast, although it is possible to compute $\refined K$ by direct evaluation of Eq.~\eqref{eq:modified kinetic energy}, this is a complex and expensive task.
Fortunately, we can estimate it rather easily by following Eastwood \textit{et al}. \cite{Eastwood_2010}, who employed numerical differentiation to estimate the time-derivatives $\dot{\vt r}$ and $\dot{\vt q}$ directly from the obtained NVE trajectory.
Considering $k$-th order estimators $\dot{\vt r}^{[k]}$ and $\dot{\vt q}^{[k]}$, we substitute Eq.~\eqref{eq:shadow ODE system} into Eq.~\eqref{eq:modified kinetic energy} to find out that
\begin{equation}
\label{eq:modified kinetic energy estimator}
\refined K = \frac{1}{2} \left( \tr{\vt p} \dot{\vt r}^{[k]} + \tr{\vt \pi} \dot{\vt q}^{[k]} \right) + \mathcal{O}(\timestep^k).
\end{equation}

For reasons that will become clear shortly, we employ an asymmetric, four-point stencil formula for estimating $\dot{\vt r}$ at a given instant $t$, which is
\begin{equation*}
\dot{\vt r}^{[3]}_t = \frac{{\vt r}_{t-2\timestep} - 6 {\vt r}_{t-\timestep} + 3 {\vt r}_t + 2 {\vt r}_{t+\timestep}}{6\timestep}.
\end{equation*}

In the case of quaternions, as a simple polynomial interpolation is insufficient to ensure that $\tr{\vt q}_i \dot{\vt q}_i = 0$ at all times for each body $i$, such as required for preserving the unit-norm constraint \cite{Silveira_2017}, we estimate the time-derivative $\dot{\vt q}$ by means of \cite{Schay_1995}
\begin{equation*}
\dot{\vt q}^{[3]}_t = {\mt \Pi}({\vt q}_t) \frac{{\vt q}_{t-2\timestep} - 6 {\vt q}_{t-\timestep} + 3 {\vt q}_t + 2 {\vt q}_{t+\timestep}}{6\timestep},
\end{equation*}
where ${\mt \Pi}(\vt q)$ is a block-diagonal projection matrix, with each diagonal block given by ${\mt \Pi}_i = {\mt 1} - {\vt q}_i \tr{\vt q}_i$, where $\mt 1$ is the identity matrix.

The observation that $\refined{\Ham{}}$ still has a quadratic-form dependency on the momenta has an important consequence: it ensures
the validity of Tolman's generalized equipartition theorem \cite{Tolman_1918, Uline_2008, Eastwood_2010} regarding a system with this Hamiltonian.
Moreover, as such dependency lies exclusively in $\refined K$, an outcome of this theorem for a system with $L$ degrees of freedom is that
\begin{equation}
\label{eq:equipartition}
L k_\mini{B} T = \left\langle \tr{\vt p} \diff{\refined K}{\vt p} + \tr{\vt \pi} \diff{\refined K}{\vt \pi} \right\rangle_\timestep = 2\langle \refined{K} \rangle_\timestep,
\end{equation}
where subscript $\timestep$ points out that this average depends on the employed time-step size.
The expression above extends the main result of Eastwood \textit{et al}. \cite{Eastwood_2010} to systems containing rigid bodies.
It provides a temperature estimator in substitution to the ordinary one, which is proportional to $\langle K \rangle$.
We emphasize that the temperature in Eq.~\eqref{eq:equipartition} concerns the system that has actually been simulated, rather than the one initially specified.

\subsection{Refinement of the NVT Dynamics}
\label{sec:refined_method}

Bringing the attention back to the NVT case, our proposal for dealing with discretization effects consists in acknowledging that the middle propagator in Eq.~\eqref{eq:splitting NVT}, if split according to Eq.~\eqref{eq:splitting NVE}, will produce a phase-space move that is more closely consistent with the modified Hamiltonian equations than with the original ones.
Thus, we should adjust the Nos\'e-Hoover chain with the aim of obtaining a distribution proportional to $e^{-\beta \refined{\Ham{}}}$ rather than $e^{-\beta \Ham{}}$, where $\beta = \frac{1}{k_\mini{B} T}$.
Fortunately, this entails solving Eq.~\eqref{eq:ODE system for NVT} almost exactly, by means of splitting, as described before.
The only required modification is that $\refined K$ should replace $K$ in Eq.~\eqref{eq:generalized force NVT}, which then becomes
\begin{equation}
\label{eq:thermostat_force}
G_1 = 2 \refined K - L k_\mini{B} T.
\end{equation}

Also, the \textit{refined extended energy} preserved by this approach is now given by
\begin{equation}
\label{eq:this extended energy}
{\refined H} = \refined K + \refined U + {\textstyle\sum\limits_{j=1}^{M}} \frac{p_{\eta_j}^2}{2Q_j} + L k_\mini{B} T\eta_1 + k_\mini{B} T {\textstyle\sum\limits_{j=2}^{M}} \eta_j.
\end{equation}

\begin{figure}
	\includegraphics{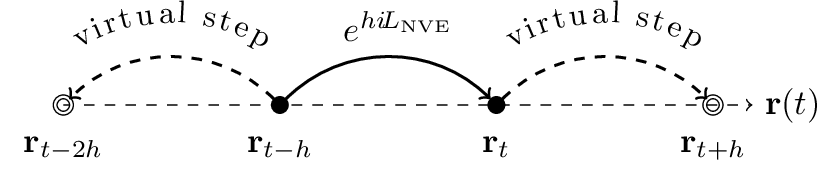}
	\caption{Schematic representation of the executed steps in a single time-step of our refined NVT dynamics. The solid arrow represents the NVE part of the integration, while the dotted arrows emphasize that forward and backward steps are carried out for positions and orientations with the only purpose of approximating their time-derivatives at point $t$.}
	\label{fig:virtual steps}
\end{figure}

One instant in which $\refined K$ must be evaluated is immediately after the action of propagator $e^{\timestep \Liu{\tiny NVE}}$.
This takes place once per time step, intercalating with non-Hamiltonian propagations.
Therefore, at such instant there is no previous record of a Hamiltonian trajectory from which we can estimate $\dot{\vt r}$ and $\dot{\vt q}$ with high accuracy.
Fortunately, only positions and orientations are required for the numerical differentiation.
This allows us to generate a four-point stencil by executing two virtual, incomplete Verlet steps, backward and forward in time, such as illustrated in Fig.~\ref{fig:virtual steps}.
By virtual and incomplete we mean, respectively, that these steps are never incorporated to the trajectory and that they are aborted once the positions and orientations have been updated.
Thus, we can obtain the estimates $\dot{\vt r}^{[3]}$ and $\dot{\vt q}^{[3]}$ without any additional force evaluation, which results in a small computational overhead.

In addition, $\refined K$ must be reevaluated after the action of propagator $e^{\frac{\timestep}{2n} {\mathcal L}_0}$, which scales both $\vt p$ and $\vt \pi$ by a factor $e^{-\alpha_1^\ast \frac{\timestep}{2n}}$.
Because such action leaves the positions and quaternions unchanged and, as a consequence, the matrix $\refined{\mathbf \Omega}(\vt r, \vt q, \timestep^2)$ in Eq.~\eqref{eq:modified kinetic energy} remains unaltered, the reevaluation of $\refined K$ is done as simply as ${\refined K} = e^{-\alpha_1^\ast \frac{\timestep}{n}} {\refined K}^\ast$.

As an example of application in stochastic thermostats, we consider the  velocity-rescaling scheme introduced by Bussi \textit{et al}. \cite{Bussi_2007}.
Its refined version can be simply obtained by substituting $K$ by $\refined{K}$ in Eq.~(A7) of Ref.~\citenum{Bussi_2007}.
In the implementation used to obtain some of the results in Sec.~\ref{sec:discretization effects}, we also split the stochastic propagator in order to have a Trotter-Suzuki type of integration (whereas a leap-frog type is used originally).

Finally, we recall that configurations are sampled from a distribution proportional to $e^{-\beta \refined{\Ham{}}}$.
Notwithstanding, it is possible to compute ensemble averages consistent with a distribution proportional to $e^{-\beta \Ham{}}$, which would have been obtained if $\timestep \to 0$.
For this, we simply need to reweight any computed observable $A$ in accordance with \cite{Torrie_1977}
\begin{equation}
\label{eq:reweighting}
\langle A \rangle_0 = \frac{\left\langle A \exp \left({\frac{\refined{\Ham{}} - \Ham{}}{k_\mini{B} T}}\right)\right\rangle_\timestep}{\left\langle \exp \left({\frac{\refined{\Ham{}} - \Ham{}}{k_\mini{B} T}}\right)\right\rangle_\timestep}.
\end{equation}

A drawback of our approach is that the resulting integrator lacks time-reversal symmetry due to the asymmetric differentiation formulas.
As a matter of fact, long-term reversibility is not generally achievable with software that relies on floating-point arithmetics.
In any event, if short-term reversibility is a requirement, one can simply employ a symmetric five-point stencil formula, at the price of one additional force computation round per time step.
Higher-order derivative estimates can also be obtained in this fashion if necessary.

\section{Application to the Simulation of Liquid Water}
\label{sec:numerical_results}

In this section, we analyze the performance of our proposed NVT-MD method and compare its results to those obtained by employing both the method of Kamberaj, Low, and Neal \cite{Kamberaj_2005} (KLN) and the NHC integrator of Martyna \textit{et al}. \cite{Martyna_1996} with our adaptations for rigid bodies.

The system under study has periodic boundary conditions and contains 903 TIP3P \cite{Jorgensen_1983} water molecules at a density of 970~kg/m\textsuperscript{3}.
The 12-6 Lennard-Jones and damped Coulombic interactions were truncated at 10 {\AA} and smoothed by a switching function over the range from 9.5 {\AA} to 10 {\AA}, such as done in Ref.~\citenum{Silveira_2017}. 
The damping of Coulombic interactions was done by a factor $\text{erfc}(-\alpha r)$, with $\alpha = 0.29~\text{\AA}^{-1}$.
In addition, all thermostated simulations were carried out considering a time-scale constant $t_d = 100~\text{fs}$ (or $\tau = 100~\text{fs}$, in the terminology of Ref.~\citenum{Bussi_2007}).

In order to verify if the total energies sampled in the simulations were consistent with the canonical ensemble, we employed the testing procedure developed by Shirts in Ref.~\citenum{Shirts_2013}.
All the NVT-MD methods investigated here succeeded in the test for time-step sizes up to 5 fs (results not shown).

\subsection{Refined Hamiltonian Calculation in NVE Simulations}

\begin{figure}
	\includegraphics{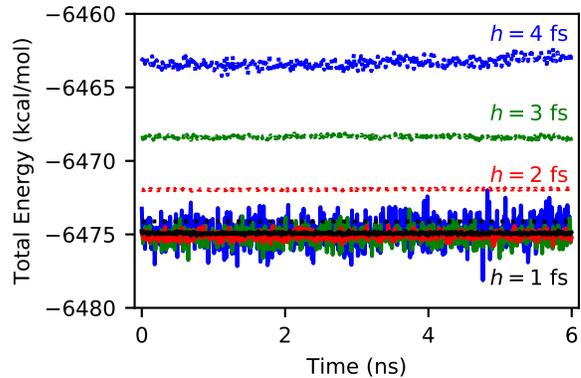}
	\caption{Specified Hamiltonian $\Ham{}$ (solid lines) and refined Hamiltonian $\refined{\Ham{}}$ (dotted lines) for different time-step sizes ($\timestep$) obtained from NVE MD simulations of 903 TIP3P \cite{Jorgensen_1983} water molecules using the numerical scheme of Eq.~\eqref{eq:splitting NVE} with the unsplit rotation method \cite{Silveira_2017}.}
	\label{fig:nve}
\end{figure}

We start by examining how much the refined Hamiltonian departs from the specified one in NVE simulations.
This is a direct way of determining the magnitude of the discretization errors \cite{Engle_2005} and is essential for discussing the NVT case.
In Fig.~\ref{fig:nve} we show the time evolution of both $\refined{\Ham{}}$ (dotted lines) and $\Ham{}$ (solid lines) for the water system using different time-step sizes.
For this, we used the numerical scheme of Eq.~\eqref{eq:splitting NVE} with the unsplit solution of Ref.~\citenum{Silveira_2017} applied for the rotations.
The four-point differentiation formulas presented in Sec.~\ref{sec:refined_method} were employed for computing the refined kinetic energy.
As expected, $\refined{\Ham{}}$ is very well conserved and is close to $\Ham{}$ when $\timestep = 1~\text{fs}$, but rapidly departs from $\Ham{}$ when $\timestep$ increases.
It is worth remarking that $\refined{\Ham{}}$ corresponds to a simple and, most importantly, computationally inexpensive approximation of the truly conserved shadow Hamiltonian.
As a result, we observe increasing fluctuations in $\refined{\Ham{}}$ for $\timestep = 3~\text{fs}$ and $\timestep = 4~\text{fs}$.
This is an indication that
1) the employed four-point formulas do not provide derivatives with the required accuracy and/or
2) the neglected high-order terms of the Baker-Campbell-Hausdorff series (see Appendix \ref{sec:rigid body shadow hamiltonian}) would contribute notably to the shadow Hamiltonian.

\subsection{Numerical Stability}
\label{sec:numerical stability}

As described in Sec.~\ref{sec:refined_method}, it is possible to introduce the refined kinetic energy $\refined K$ in an NVT integrator without a substantial increase in computational effort.
However, this can only be achieved by limiting the order of approximation of numerical derivatives and sacrificing short-term reversibility.
This introduces the question of whether these conditions will influence the long-term stability of the method, especially if one tries to make use of large time steps.

\begin{figure}
	\includegraphics{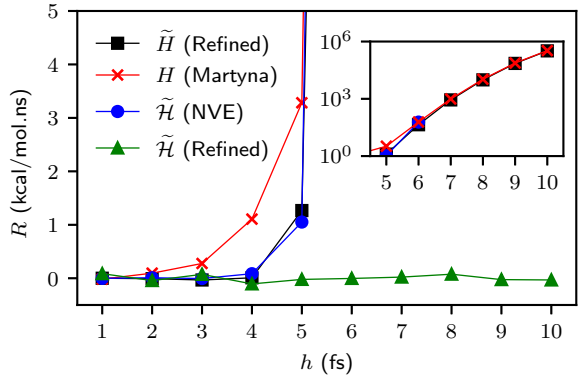}
	\caption{Influence of the time-step size ($\timestep$) on the long-term drift rate ($R$) in: the extended energy ($\refined{H}$, black squares) and its refined Hamiltonian ($\refined{\Ham{}}$, green up triangles) part for our NVT method, as well as in the extended energy ($H$, red x) for the method due to Martyna \textit{et al}. \cite{Martyna_1996} and in the refined Hamiltonian ($\refined{\Ham{}}$, blue circles) for the numerical scheme of Eq.~\eqref{eq:splitting NVE} with unsplit solution of rotation.}
	\label{fig:energy_drift}
\end{figure}

Here we define a drift rate, denoted by $R$, as the long-term rate of change of a quantity in a supposedly equilibrated MD simulation.
It is defined as the slope of a least-square regression line expressing such quantity as a function of time.
Recall that the symmetric scheme of Martyna \textit{et al}. \cite{Martyna_1996} is supposed to preserve the extended energy $H$ given by Eq.~\eqref{eq:nvt extended energy}.
In the case of our proposed method, which we refer to as Refined NHC, the conserved extended energy is $\refined H$, such as defined by Eq.~\eqref{eq:this extended energy}.
In Fig.~\ref{fig:energy_drift}, we present drift rates of these quantities computed in liquid water simulations, with $18~\text{ns}$ of production time, carried out with different time-step sizes.
As one can observe, long-term stability is slightly improved by our method if $\timestep \le 5~\text{fs}$, indicating that there is little or no impact of the lack of time-reversal symmetry on $R$.

Fig.~\ref{fig:energy_drift} also contains drift rates computed for the refined Hamiltonian $\refined{\Ham{}}$ in an NVE simulation \cite{Silveira_2017}.
Up to $h = 6~\text{fs}$, these rates practically coincide with those obtained for $\refined H$ with the Refined NHC scheme, indicating that the major causes for the observed drift takes place during evaluations of the NVE propagator.
As the interaction potential we use is very smooth at cutoff, a possible cause is the amplification of round-off errors which can occur in events of close interatomic approaches \cite{Engle_2005}.
For time steps larger than $6~\text{fs}$, the NVE integration becomes fully unstable.
In contrast, all NVT integrators manage to remain numerically stable at least up to $h = 10~\text{fs}$, but at the cost of displaying exceedingly large drifts in their corresponding extended energies, as one can see the inset of Fig.~\ref{fig:energy_drift}.
Trying to look for the cause of this behavior, we also present drift rates computed for the refined Hamiltonian $\refined{\Ham{}}$ in the simulations carried out with the Refined NHC method.
Recall that this part of $\refined H$ represents the total energy of the system.
Note that no drift occurs for $\refined{\Ham{}}$, not even for the largest considered time steps, showing that the thermostat stabilizes the numerical integration of the physical variables.
This has already been noticed by Davidchack \cite{Davidchack_2010} and also by Bond and Leimkuhler \cite{Bond_2007}, who in this respect stated that ``the thermostat acts as a sort of reservoir for numerical errors''.

\begin{figure}
	\includegraphics{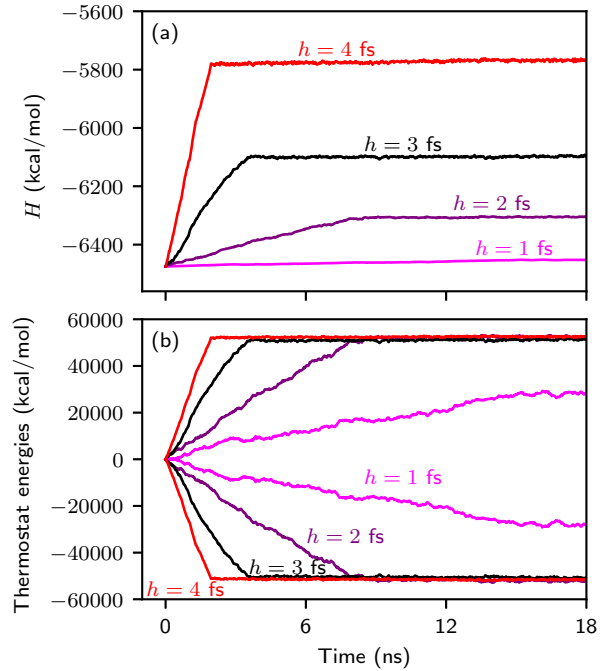}
	\caption{Time evolution of (a) the extended energy and (b) of each thermostat energy for the NVT method introduced by Kamberaj, Low, and Neal \cite{Kamberaj_2005}. In part (b), the negative values correspond to the thermostat coupled to the translational degrees of freedom.}
	\label{fig:num_stab}
\end{figure}

To end this section, we report in Fig.~\ref{fig:num_stab} an anomalous behavior observed when the KLN method \cite{Kamberaj_2005} is employed.
Particularly, one can see in Fig.~\ref{fig:num_stab}(a) that, for each time-step size, the extended energy $H$ drifts upwards until reaching a steady-state value at which the integrator becomes stable.
These results were obtained with our own simulation code and double-checked using LAMMPS \cite{Plimpton_1995} with a customized pair style.
As already mentioned, the KLN scheme \cite{Kamberaj_2005} couples separate NHC thermostats to the rotational and translational degrees of freedom.
Considering a different factorization scheme, Davidchack \cite{Davidchack_2010} discovered the existence of energy flows between the two thermostats.
This is confirmed for the KLN method \cite{Kamberaj_2005} in Fig.~\ref{fig:num_stab}(b), where we depict separately the contribution of each thermostat to the extended energy $H$.
In this figure, the negative energy values always correspond to the thermostat coupled to the translational degrees of freedom.
Therefore, energy flows from this thermostat to the one coupled to the rotational degrees.
Interestingly, Fig.~\ref{fig:num_stab} shows that the rate at which the thermostats approach steady state decreases with $\timestep$. As far as we know, this behavior has not been reported before.

\subsection{Correction of Discretization Effects on Computed Thermodynamic Properties}
\label{sec:discretization effects}

In the previous section, we have seen that drifts in the extended energy of an NVT simulation can accumulate in the thermostat-related variables, thus preventing numerical instability from happening.
This allows simulations to be executed with large time steps, which could in principle be used to save computer time.
However, as we will show in this section, discretization errors can become manifest before any risk of instability arises.
Fortunately, our proposed methodology is capable of mitigating the effect of these errors on ensemble averages computed for the simulated system.

\begin{figure*}
	\includegraphics{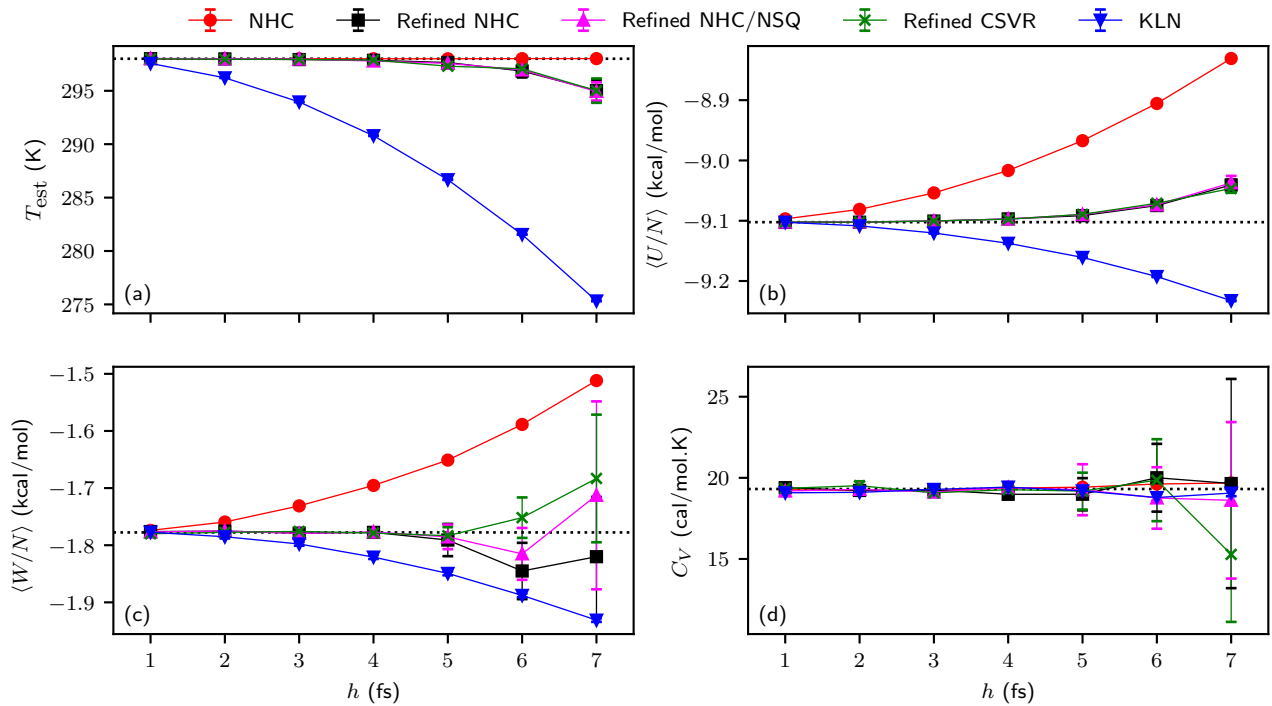}
	\caption{Effect of the time-step size on (a) the estimated temperature, (b) the mean potential energy per molecule, (c) the mean internal virial per molecule, and (d) the constant-volume specific heat of water, as obtained from NVT-MD simulations of 903 TIP3P \cite{Jorgensen_1983} water molecules at $T=298~\mathrm{K}$ and $\rho=970~\mathrm{kg/m^3}$ and employing different numerical integration schemes.
	The NHC scheme ($\bullet$) corresponds to the Nos\'{e}-Hoover Chain integrator of Martyna \textit{et al}. \cite{Martyna_1996} extended to rigid bodies with analytically computed free rotations \cite{Silveira_2017}.
	The Refined NHC scheme ($\blacksquare$) is obtained by subjecting the preceding method to the refinement procedure described in Sec.~\ref{sec:refined_method}.
	The Refined NHC/NSQ scheme ($\blacktriangle$) differs from Refined NHC only in that the approximate NO-SQUISH \cite{Dullweber_1997, Miller_2002} solution is used for free rotations.
	The Refined CSVR scheme ($\times$) corresponds to the canonical stochastic velocity rescaling method \cite{Bussi_2007} subject to refinement as well.
	Finally, the KLN scheme ($\blacktriangledown$) is the double-thermostat method of Kamberaj, Low, and Neal \cite{Kamberaj_2005}.}
	\label{fig:properties}
\end{figure*}

In Fig.~\ref{fig:properties}, we present some thermodynamic properties computed for our liquid-water system using several NVT-MD methods and time-step sizes.
All simulations were carried out for $T = 298~\text{K}$ and the analyzed quantities are the estimated temperature
\begin{equation*}
T_\text{est} = \frac{2\langle K \rangle}{(6N-3)k_\mini{B}},
\end{equation*}
the mean potential energy per molecule $\langle U/N \rangle$,
the mean internal virial per molecule $\langle W/N \rangle$, and
the constant-volume specific heat capacity
\begin{equation*}
c_\mini{V} = \frac{\partial \langle \Ham{}/N \rangle}{\partial T} = \frac{\langle \Ham{}^2 \rangle - \langle \Ham{} \rangle^2}{N k_\mini{B} T^2}.
\end{equation*}

The internal virial $W$ is computed as described in Ref.~\citenum{Silveira_2017}.
All graphical symbols in Fig.~\ref{fig:properties} represent results obtained after $18~\text{ns}$ of production time, while the lines are just guides for the eyes.
The horizontal dotted lines are meant to help us visualize the departure of the computed properties from their expected values, which are estimated by averaging the results obtained from the three refined methods (see details below) with $\timestep = 1~\text{fs}$.
Error bars represent 95\% confidence intervals estimated via overlapping batch means \cite{Meketon_1984, Flegal_2010} and standard uncertainty propagation formulas.

Fig.~\ref{fig:properties} contains results of five different NVT integrators.
Our measure-preserving extension to rigid bodies of the splitting scheme by Martyna \textit{et al}. \cite{Martyna_1996} is the basis of three of them.
Recall that this is a deterministic method in which a single Nos\'{e}-Hoover chain \cite{Martyna_1992} is attached to all degrees of freedom.
First, the method is applied in a conventional (i.e. unrefined) way with the unsplit solution of Ref.~\citenum{Silveira_2017} being used for handling rotations.
Then, it is modified by introducing the refinement/reweighting procedure proposed in Sec.~\ref{sec:refined_method}.
In Fig.~\ref{fig:properties}, these two integrators are referred to as NHC and Refined NHC, respectively.
Next, we replace the unsplit rotation solution by the simpler NO-SQUISH splitting of Miller \textit{et al}. \cite{Miller_2002}, denoting the resulting integrator by Refined NHC/NSQ in Fig.~\ref{fig:properties}.
We note that our refinement procedure is only rigorously valid (see Appendix \ref{sec:rigid body shadow hamiltonian}) for unsplit rotations.
Nevertheless, it is known that NO-SQUISH rotations approach them almost indistinguishably in the case of water if small time steps are employed \cite{Silveira_2017}.
The fourth integrator we test is a refined version of the velocity rescaling method of Bussi, Donadio, and Parrinello \cite{Bussi_2007}, which differs from the others by involving a global stochastic thermostat.
This is identified in Fig.~\ref{fig:properties} as Refined CSVR (acronym for Canonical Stochastic Velocity Rescaling).
Finally, we also consider the double-thermostat KLN integrator \cite{Kamberaj_2005} for comparison purposes, despite the spurious energy flux it generates (see Sec.~\ref{sec:numerical stability}).
This is to show that the method, denoted by KLN in Fig.~\ref{fig:properties}, also exhibits discretization-related issues not present in other integrators.
We stress that thermodynamic properties are estimated in distinct ways for the conventional and the refined integrators.
For the former class, simple averages are taken from the sampled configurations.
For the latter, reweighting is carried out via Eq.~\eqref{eq:reweighting}.

In Fig.~\ref{fig:properties}(a), we see that the estimated temperature for the thermostat identified as KLN decreases appreciably as the time step size increases.
This behavior is typical of Verlet-like NVE integrators\cite{Davidchack_2010, Silveira_2017}, indicating that the KLN factorization scheme \cite{Kamberaj_2005} produces a relatively weak effect on the particle momenta.
On the other hand, the stringent temperature control achieved by the NHC scheme leads to a severely degraded accuracy in both $\langle U/N \rangle$ and $\langle W/N \rangle$, as noted in Figs.~\ref{fig:properties}(b) and \ref{fig:properties}(c), respectively.
According to the results of Fig.~\ref{fig:nve}, discretization errors in the NVE integration become manifest at $2~\text{fs}$.
Thus, the apparent success of the unrefined NHC scheme, in terms of kinetic energy control along the entire range of time-step sizes considered, is wiped out by those discretization errors.
If not handled properly, they can yield unreliable configurational properties.

From Fig.~\ref{fig:properties}, it becomes clear that the refined schemes produce significantly improved results. For $\timestep \leq 5~\mathrm{fs}$, we are able to accurately reproduce the specified temperature and compute other thermodynamic properties considered.
For longer time steps, a small but clear systematic error is present, as the computed averages slightly depart from the reference values.
In principle, this could be mitigated by using a higher-order approximation to the shadow Hamiltonian.
Nonetheless, we cannot anticipate whether the ability to increase the step size would compensate the additional effort required per time step.

As stated above, the NO-SQUISH solution \cite{Dullweber_1997, Miller_2002} is very accurate for the system and time-step sizes considered here \cite{Silveira_2017},
which explains the remarkable agreement between the results shown in Fig.~\ref{fig:properties} for the Refined NHC and Refined NHC/NSQ schemes.
This is an advantage because the NO-SQUISH method is much simpler to implement than the exact rotation solution and is already available in LAMMPS \cite{Plimpton_1995} and other MD packages.

Regarding the refinement of stochastic methods, the results obtained from the Refined CSVR scheme are similar to those from other refined ones.
This was expected because the unrefined versions of the CSVR \cite{Bussi_2007} (stochastic) and the NHC \cite{Martyna_1996} (deterministic) thermostats for rigid bodies produce virtually identical properties as functions of the time-step size (results not shown).
Similar behavior is expected for other global thermostats such as the Nos\'{e}-Hoover-Langevin method \cite{Samoletov_2007, Leimkuhler_2009}.
Extension of our approach to massive Langevin-type thermostats \cite{Brunger_1984, Davidchack_2009, Davidchack_2015}, in which every degree of freedom is acted upon independently, is beyond the scope of this work.
Nonetheless, we envision that this would entail using the components of entrywise products between $\vt p$ and $\dot{\vt r}^{[n]}$, and also between ${\vt \pi}$ and $\dot{\vt q}^{[n]}$ in the case of rigid bodies, instead of the scalar products present in Eq.~\eqref{eq:modified kinetic energy estimator}.

Fig.~\ref{fig:properties}(d) contains results for the constant-volume specific heat capacity, which
are not noticeably influenced by the discretization errors when $\timestep \leq 5~\mathrm{fs}$, not even if the KLN method is employed.
This means that the numerical artifacts do not affect the variance as much as they affect the mean of the total energy distribution.

\begin{figure}
	\includegraphics{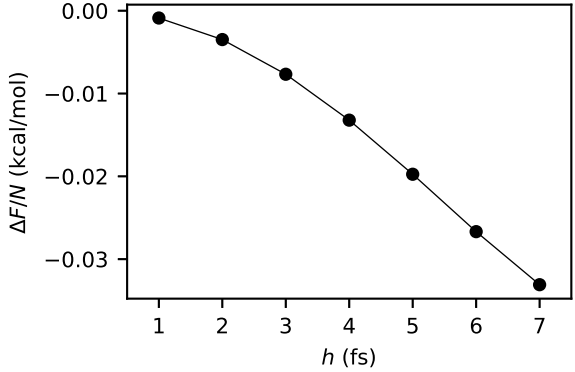}
	\caption{Difference in free energy per water molecule between the actually sampled and the target states as a function of the time-step size. The sampling was done by employing the Refined NHC scheme, while the target state is consistent with a distribution proportional to $e^{-\beta \Ham{}}$, where $\Ham{}$ corresponds to the specified Hamiltonian.}
	\label{fig:free_energy}
\end{figure}

With the aim of highlighting the importance of the reweighting procedure, in Fig.~\ref{fig:free_energy} we show the free energy difference per water molecule ($\Delta F/N$) between the actually sampled and the target states, obtained with the Refined NHC scheme as a function of the time-step size.
Note that $\Delta F$ corresponds to the logarithm of the denominator of Eq.~\eqref{eq:reweighting}.
It should be mentioned that the other refined schemes yield similar results and, as expected, the free energy difference is non-negligible except for the smallest time steps.

\begin{figure*}
	\includegraphics{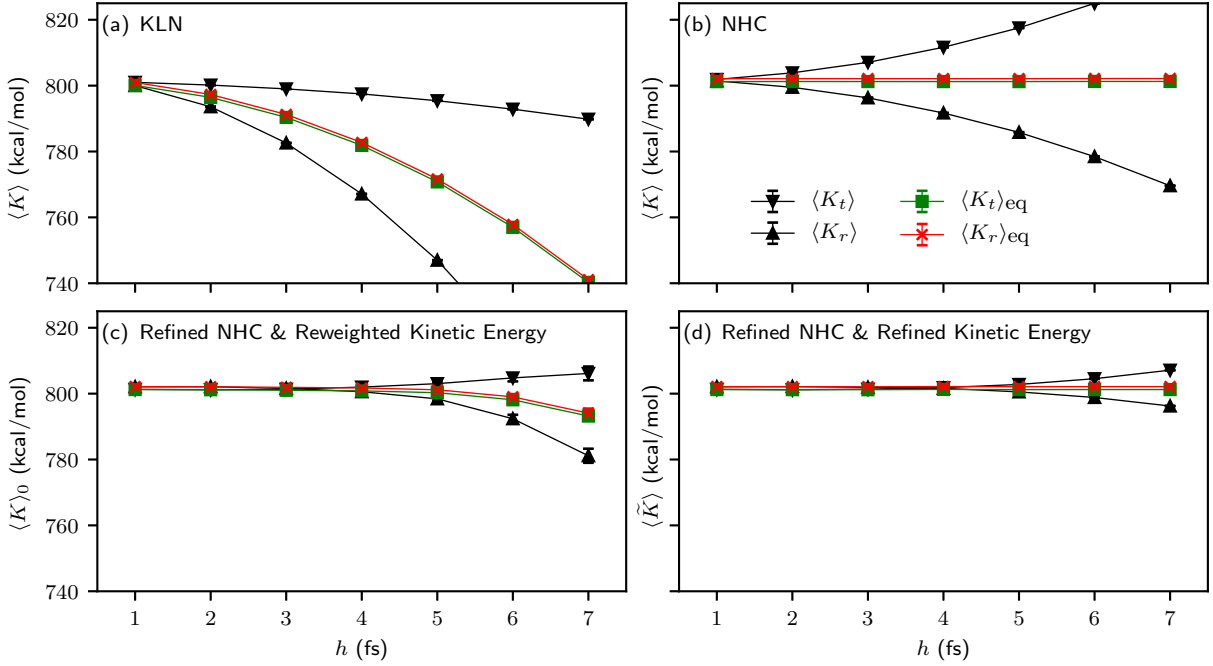}
		\caption{Influence of the time-step size on the kinetic energy partition in NVT-MD simulations of 903 TIP3P \cite{Jorgensen_1983} water molecules.
		Part (a) contains results obtained from the KLN method \cite{Kamberaj_2005}.
		The results in Part (b) were obtained by using the NHC method \cite{Martyna_1996} extended for rigid bodies and without any refinement.
		In Part (c), they  were obtained from our Refined NHC \cite{Martyna_1996} method, with the kinetic energies computed at the target state via reweighting.
		Finally, Part (d) depict the refined kinetic energies at every sampled state resulting from the same simulations of Part (c).
		In all parts, we present the simulated translational ($\blacktriangledown$) and rotational ($\blacktriangle$) kinetic energies, as well as the expected translational ($\blacksquare$) and rotational ($\times$) values assuming equipartition.}
	\label{fig:energy_partition}
\end{figure*}

We end this section with a discussion about discretization effects on the kinetic energy partition amongst translational and rotational degrees of freedom.
Davidchack \cite{Davidchack_2010} has already performed a thorough analysis of such effects in thermostated rigid-body dynamics.
Here we complement his study by including the KLN method \cite{Kamberaj_2005} and by showing that the refinement procedure is able to mitigate the observed artifacts.
In Fig.~\ref{fig:energy_partition}, we present computed averages of the translational ($K_t$) and rotational ($K_r$) kinetic energies as functions of the time-step size, and compare them to the values that would be expected by assuming equipartition.
In Figs.~\ref{fig:energy_partition}(a) to \ref{fig:energy_partition}(c), these were calculated from the average kinetic energy $\langle K \rangle$ by means of $\langle K_t \rangle_\text{eq}$ = $\frac{3N-3}{6N-3} \langle K \rangle$ and $\langle K_r \rangle_\text{eq}$ = $\frac{3N}{6N-3} \langle K \rangle$.
For Fig.~\ref{fig:energy_partition}(d), $K$ is simply replaced by $\refined{K}$ in these equations.
In a previous work \cite{Silveira_2017}, we have shown that the rotational energy is more strongly influenced by the time-step size in the NVE case than the translational energy.
This occurs because the rotational degrees of freedom display the fastest motion in a rigid water molecule \cite{Silveira_2017}.
In Fig.~\ref{fig:energy_partition}(a), one can see that the KLN method \cite{Kamberaj_2005} tends, once again, to reproduce the behavior of the NVE integration.
The results of Fig.~\ref{fig:energy_partition}(b) were obtained by using the NHC scheme of Martyna \textit{et al}. \cite{Martyna_1996} applied to rigid bodies.
These results are similar to those shown in Fig.~\ref{fig:energy_partition}(a), in the sense that the simulated translational and rotational components deviate upwards and downwards, respectively, from their equipartioned values.
This observation is in consonance with the analysis of Ref.~\citenum{Davidchack_2010}.
To some extent, our refinement/reweighting procedure is able to correct these deviations, as one can see in Fig.~\ref{fig:energy_partition}(c).
Finally, by observing Fig.~\ref{fig:energy_partition}(d), one can note that the refined translational and rotational kinetic energies, 
$\refined K_t = \frac{1}{2} \tr{\vt p} \dot{\vt r}^{[3]}$ and $\refined K_r = \frac{1}{2} \tr{\vt \pi} \dot{\vt q}^{[3]}$, being estimates of their ``shadow'' counterparts, comply with Tolman's equipartition theorem \cite{Tolman_1918} for time-step sizes up to $5~\text{fs}$.

\section{Concluding Remarks}
\label{sec:conclusion}

In this paper, we have devised a simple, general approach that integrates backward error analysis into Nos\'{e}-Hoover chains and other global thermostats.
The rationale for our proposal arises from recognizing that a ``shadow kinetic energy'' should substitute the conventional one in the equations that dictate the thermostat action.
In practice, we employ backward error analysis to compute, with small computational overhead, a refined version of the kinetic energy, which serves as an estimate for its shadow analogue.
This turns out to be crucial for simulations that display considerable discretization errors, in which a more precise definition of the simulation temperature \cite{Eastwood_2010} is obtained by applying Tolman's generalized equipartition theorem to the refined kinetic energy rather than to the conventional one.

Significant enhancement was achieved for a system of rigid TIP3P-water molecules \cite{Jorgensen_1983}.
Analytical derivation of a refined Hamiltonian formula for rigid bodies was greatly facilitated by considering an unsplit integration of rotations, such as in the method we have recently introduced \cite{Silveira_2017}.
In practice, we have been able to remove serious artifacts which take place when the method of Kamberaj, Low, and Neal \cite{Kamberaj_2005} or a rigid-body extension of the standard Nos\'{e}-Hoover chain thermostat \cite{Martyna_1992, Martyna_1996} is used with time steps larger than $1~\text{fs}$.
The refinement/reweighting approach described in Sec.~\ref{sec:refined_method} corrected discretization effects in computed thermodynamic properties of water for time-step sizes up to $5~\text{fs}$. This is of paramount importance as it allows us to actually exploit the useful coarse-graining strategy provided by rigid-body MD by increasing the time-step size without sacrificing accuracy.

Finally, it is worth remarking that our method is not limited to rigid bodies.
It can also be employed in conjunction with other constrained MD schemes, such as SHAKE \cite{Ryckaert_1977} and RATTLE \cite{Andersen_1983}, for instance.
Of course, constraints must be properly taken into account in the computation of time-derivatives of atomic positions, as was done here for the time-derivatives of quaternions.

\begin{acknowledgements}
This work is dedicated to the memory of Prof. Affonso da Silva Telles, a true scientist who played a major role in laying the foundation of Chemical Engineering research in Brazil.

The authors acknowledge the financial support provided by Petrobras (project code CENPES 16113).
\end{acknowledgements}

\appendix

\section{Shadow Hamiltonian Approximation for Rigid Bodies}
\label{sec:rigid body shadow hamiltonian}

For two non-commutative Liouville operators $\Liu A$ and $\Liu B$ and a real constant $\timestep$, the symmetric Baker-Campbell-Hausdorff (BCH) formula can be expressed as \cite{Hairer_2006}
\begin{equation}
\label{eq:symmetric BCH}
\begin{split}
&e^{\frac{\timestep}{2} \Liu B} e^{\timestep \Liu A} e^{\frac{\timestep}{2} \Liu B} = \\
&= e^{\timestep (\Liu A + \Liu B) + \frac{\timestep^3}{24} \left[2 \Liu A + \Liu B,[\Liu A,\Liu B]\right] + \mathcal{O}(\timestep^5)},
\end{split}
\end{equation}
where $[X,Y] = XY - YX$ is the commutator of operators $X$ and $Y$.
The remaining terms of the infinite series involve ever-deepening nested commutators \cite{Hairer_2006}.

Let us now represent the operators $\Liu A$ and $\Liu B$ using Poisson brackets involving Hamiltonians $\Ham A$ and $\Ham B$, that is, $\Liu A = \{\circ,\Ham A\}$ and $\Liu B = \{\circ,\Ham B\}$.
By employing the property of anti-commutativity and the Jacobi identity \cite{Hairer_2006}, we can deduce that
\begin{align*}
[\Liu A,\Liu B] &= \{\{\circ,\Ham B\},\Ham A\} - \{\{\circ,\Ham A\},\Ham B\} = \\
&= -\{\Ham A,\{\circ,\Ham B\}\} - \{\Ham B,\{\Ham A,\circ\}\} = \\
&= \{\circ,\{\Ham B,\Ham A\}\} = \\
&= \{\circ,{\Liu A} {\Ham B}\}.
\end{align*}

This means that the commutator $[\Liu A,\Liu B]$ is, in fact, a new Liouville operator $\Liu C$ whose associated Hamiltonian is $\Ham C = {\Liu A}{\Ham B}$.
By applying this procedure recursively, we can rewrite the right-hand side of Eq.~\eqref{eq:symmetric BCH} as $e^{\timestep \Liu S}$, where $\Liu S$ is the Liouville operator whose associated Hamiltonian is
\begin{equation*}
\label{eq:general shadow hamiltonian}
\Ham{S} = \Ham A + \Ham B + \frac{\timestep^2}{24} (2 \Liu A + \Liu B){\Liu A}{\Ham B} + \mathcal{O}(\timestep^4).
\end{equation*}

Therefore, a splitting method meant to approximately reproduce the dynamics of a system with Hamiltonian $\Ham{} = \Ham A + \Ham B$ will, in fact, reproduce exactly (round-off issues aside) the dynamics of another system with Hamiltonian $\Ham{S}$ as above.
In the literature, $\Ham{S}$ is usually referred to as a shadow Hamiltonian.

We remark that a single rigid body is considered in the development that follows, but the final result can be readily generalized for a system with multiple bodies.

For the splitting introduced in Sec.~\ref{sec:nve}, $\Ham A = K(\vt p, \vt q, \vt \pi)$ and $\Ham B = U(\vt r, \vt q)$ with the corresponding Liouville operators $\Liu{A} = \tr{K_{\vt p}}\diff{}{\vt r} + \tr{K_{\vt \pi}}\diff{}{\vt q} - \tr{K_{\vt q}}\diff{}{\vt \pi}$ and $\Liu{B} = -\tr{U_{\vt r}}\diff{}{\vt p} - \tr{U_{\vt q}}\diff{}{\vt \pi}$.
Note that any gradient like $\diff{f}{\vt x}$ is now represented as $f_{\vt x}$ for the sake of simplicity.
We start by obtaining $\Liu A \Ham B = \tr{K_{\vt p}} U_{\vt r} + \tr{K_{\vt \pi}} U_{\vt q}$. Next, we deduce that
\begin{align*}
\Liu A \Liu A \Ham B &= \tr{K_{\vt p}} U_{\vt r \vt r} K_{\vt p}
+ 2 \tr{K_{\vt \pi}} U_{\vt r \vt q} K_{\vt p}
+ \tr{K_{\vt \pi}} U_{\vt q \vt q} K_{\vt \pi} \\
&+ \tr{K_{\vt \pi}} K_{\vt \pi \vt q} U_{\vt q}
- \tr{K_{\vt q}} K_{\vt \pi \vt \pi} U_{\vt q},
\end{align*}
where the fact that $\tr{U_{\vt q \vt r}} = U_{\vt r \vt q}$ has been used.
As reported in Ref.~\citenum{Silveira_2017}, the identity $\tr{\mt B}(\vt q) {\vt \pi} = -\tr{\mt B}(\vt \pi) {\vt q}$ is useful for performing differentiations.
The first-order derivatives of $K$ with respect to $\vt p$, $\vt q$, and $\vt \pi$ are, respectively,
\begin{align*}
&K_{\vt p} = {\mt M}^{-1} {\vt p}, \\
&K_{\vt q} = -\frac{1}{2} {\mt B}(\vt \pi) {\vt \omega}, \; \text{and} \\
&K_{\vt \pi} = \frac{1}{2} {\mt B}(\vt q) {\vt \omega},
\end{align*}
where $\vt \omega = \frac{1}{2} {\mt I}^{-1} \tr{\mt B}(\vt q) \vt \pi = -\frac{1}{2} {\mt I}^{-1} \tr{\mt B}(\vt \pi) \vt q$.\cite{Silveira_2017} Hence,
\begin{gather*}
\tr{K_{\vt p}} U_{\vt r \vt r} K_{\vt p} = \tr{\vt p} {\mt M}^{-1} U_{\vt r \vt r} {\mt M}^{-1} {\vt p}, \\
\tr{K_{\vt \pi}} U_{\vt r \vt q} K_{\vt p} = \frac{1}{2} \tr{\vt \omega} \tr{\mt B}(\vt q) U_{\vt r \vt q} {\mt M}^{-1} {\vt p}, \\
\tr{K_{\vt \pi}} U_{\vt q \vt q} K_{\vt \pi} = \frac{1}{4} \tr{\vt \omega} \tr{\mt B}(\vt q) U_{\vt q \vt q} {\mt B}(\vt q) \vt \omega.
\end{gather*}

As $\diff{\tr{\vt \omega}}{\vt \pi} = \frac{1}{2} {\mt B}(\vt q) {\mt I}^{-1}$, the second-order derivative of $K$ with respect to $\vt \pi$ turns out to be
\begin{equation*}
K_{\vt \pi \vt \pi} = \frac{1}{4} {\mt B}(\vt q) {\mt I}^{-1} \tr{\mt B}(\vt q).
\end{equation*}

Pre-multiplication by $\tr{K_{\vt q}}$ will introduce a product $\tr{\mt B}(\vt \pi) {\mt B}(\vt q)$.
This has been shown in Ref.~\citenum{Silveira_2017} to be equal to ${\mt S}\left( \tr{\mt B}(\vt \pi) {\vt q} \right) = -2 {\mt S}({\mt I} {\vt \omega})$, where the operator ${\mt S}(\cdot)$ builds a skew-symmetric matrix from the entries of a vector.
In addition, post-multiplication by $U_{\vt q}$ will introduce a product $\tr{\mt B}(\vt q) U_{\vt q}$, which is equal to $-2 {\vt \tau}_b$ \cite{Silveira_2017}, where ${\vt \tau}_b$ is the body-fixed frame representation of the torque.
Then, the fact that ${\mt S}(\vt x) {\vt y} = -{\mt S}(\vt y) {\vt x}$ leads to
\begin{equation*}
\tr{K_{\vt q}} K_{\vt \pi \vt \pi} U_{\vt q} = \frac{1}{2} \tr{\vt \omega} {\mt S}({\mt I}^{-1} {\vt \tau}_b) {\mt I} {\vt \omega}.
\end{equation*}

The term above is identically null, as it corresponds to the quadratic form built with a skew-symmetric matrix.
Another helpful identity taken from Appendix B of Ref.~\citenum{Silveira_2017} is ${\mt B}(\vt q){\vt \omega} = ( \sum_{j=1}^3 \omega_j \hat{\mt B}_j ) \vt q$, where each $\hat{\mt B}_j$ is a skew-symmetric permutation matrix (i.e. $\tr{\hat{\mt B}}_j = -\hat{\mt B}_j$).
This allows us to obtain
\begin{align*}
K_{\vt \pi \vt q} &= \frac{1}{2} \left(\diff{\tr{\vt \omega}}{\vt q}\right) \tr{\mt B}(\vt q) + \frac{1}{2} \sum_{j=1}^3 \omega_j \tr{\hat{\mt B}}_j = \\
&= -\frac{1}{4} {\mt B}(\vt \pi) {\mt I}^{-1} \tr{\mt B}(\vt q) - \frac{1}{2} \sum_{j=1}^3 \omega_j \hat{\mt B}_j.
\end{align*}

Pre-multiplication by $\tr{K_{\vt \pi}}$ and post-multiplication by $U_{\vt q}$, followed by some algebraic transformations, ultimately lead to
\begin{equation*}
\tr{K_{\vt \pi}} K_{\vt \pi \vt q} U_{\vt q} = -\frac{1}{2} \tr{\vt \omega} \tr{\mt S}({\mt I}^{-1} {\vt \tau}_b) {\mt I} {\vt \omega} + \frac{1}{2} \tr{\vt \omega} {\mt S}({\vt \tau}_b){\vt \omega}.
\end{equation*}

For the same reason explained above, both terms in the right-hand side of the equation above are identically null.
We are now able to evaluate $\Liu A \Liu A \Ham B $, which is
\begin{align*}
\Liu A \Liu A \Ham B = &\tr{\vt p} {\mt M}^{-1} U_{\vt r \vt r} {\mt M}^{-1} {\vt p} + \tr{\vt \omega} \tr{\mt B}(\vt q) U_{\vt r \vt q} {\mt M}^{-1} {\vt p}
+ \\
&\frac{1}{4} \tr{\vt \omega} \tr{\mt B}(\vt q) U_{\vt q \vt q} {\mt B}(\vt q) \vt \omega.
\end{align*}

Finally, the term that remains for obtaining $\Ham{S}$ is given by
\begin{align*}
\Liu B \Liu A \Ham B &= -\tr{U_{\vt r}} \tr{K_{\vt p \vt p}} U_{\vt r} - \tr{U_{\vt q}} \tr{K_{\vt \pi \vt \pi}} U_{\vt q} = \\
&= -\tr{\vt F} {\mt M}^{-1} {\vt F} - \tr{\vt \tau}_b {\mt I}^{-1} {\vt \tau}_b.
\end{align*}

It is convenient to rewrite the total kinetic energy of the original system, Eq.~\eqref{eq:kinetic energy}, as a quadratic form $K = \frac{1}{2}\tr{[\substack{\vt p \\ \vt \pi}]} {\mt \Omega}(\vt q) [\substack{\vt p \\ \vt \pi}]$, where $\mt \Omega$ is a symmetric, block-diagonal matrix defined as
\begin{equation}
\label{eq:block-diagonal inverse mass tensor}
{\mt \Omega} = \left[\begin{array}{cc}
{\mt M}^{-1} & \mt 0 \\
\mt 0 & \frac{1}{4} {\mt B}(\vt q) {\mt I}^{-1} \tr{\mt B}(\vt q)
\end{array}\right].
\end{equation}

We now proceed to present the final expression of $\Ham{S}$, which results from grouping the terms corresponding to forces and torques in the refined potential energy $\refined U(\vt r, \vt q, h^2)$, given by
\begin{equation*}
\refined U = U - \frac{\timestep^2}{24} \left( \tr{\vt F} {\mt M}^{-1} {\vt F} + \tr{\vt \tau} \tr{\mt A} {\mt I}^{-1} {\mt A} {\vt \tau} \right),
\end{equation*}
while the terms that depend on the momentum vectors $\vt p$ and $\vt q$ are grouped together in the refined kinetic energy $\refined K$, which can be written in a compact matrix form as
\begin{equation*}
\refined K = \frac{1}{2} \tr{ \left[\begin{array}{c} \vt p \\ \vt \pi \end{array}\right]} \refined{\mathbf \Omega}(\vt r, \vt q, h^2) \left[\begin{array}{c} \vt p \\ \vt \pi \end{array}\right],
\end{equation*}
where
\begin{equation*}
\refined{\mt \Omega} = {\mt \Omega} + \frac{\timestep^2}{6} {\mt \Omega} \left[\begin{array}{cc}
U_{\vt r \vt r} & U_{\vt r \vt q} \\
U_{\vt q \vt r} & U_{\vt q \vt q}
\end{array}\right] \tr{\mt \Omega}.
\end{equation*}

Finally, the shadow Hamiltonian is given by $\Ham{S} = {\refined U} + {\refined K} + {\mathcal O}(\timestep^4)$.

\bibliography{rigid_bodies}

\end{document}